УДК 338: 519.1

# Алгоритмы поиска эффективных логистических цепей

Сергей Александрович Лавренченко, доцент, lawrencenko@hotmail.com

Ирина Альбертовна Дуборкина, доцент, duborkina2012@yandex.ru

ФГБОУ ВПО «Российский государственный университет туризма и сервиса», факультет туризма и сервиса (г. Люберцы), отделение кафедры экономики и управления

**Аннотация.** Логистические сети возникают, когда имеет место передача по звеньям (каналам) материальной субстанции или материальных объектов (например, зарегистрированного багажа на авиарейсах), а также энергии, информации или финансов. В работе предлагается и обосновывается общая концепция логистический сети для моделирования сервиса любого рода, доставляемого по звеньям между узлами сети. Для каждого звена сети его эффективность определяется как отношение объема полезного сервиса на выходе звена к объему затраченного сервиса на его входе (за определенный период времени). Так же определяется эффективность цепи в сети — это отношение объема сервиса на выходе цепи к объему сервиса на ее входе. Общая эффективность цепи вычисляется как произведение эффективностей ее звеньев, и чем больше эффективность цепи, тем меньше в ней потерь. В работе вводится понятие убыточности сервиса таким образом, чтобы общая убыточность цепи равнялась сумме убыточностей ее звеньев, т.е. эффективности звеньев перемножаются, а их убыточности складываются. Таким образом, антагонистическая пара (эффективность, убыточность) оказывается аналогичной паре (надежность, энтропия) из теории связи. Представлены возможные интерпретации предлагаемой логистической модели: энергетические, материальные, информационные и финансовые сети. Предложены алгоритмы для информационного обеспечения логистических сетей: два алгоритма для отыскания максимально эффективной цепи из указанного узла отправления в указанный узел назначения и два алгоритма для отыскания в данной сети гарантированного минимального уровня эффективности сервиса между любой парой не указанных конкретно узлов. Продемонстрирован пример применения одного из алгоритмов для нахождения максимально эффективной энергетической цепи от подстанции к указанному потребителю в конкретной энергетической сети.

**Ключевые слова:** логистика, энергетическая цепь, эффективность.

**1. Введение.** Под *графом* подразумевается объект, состоящий из данного множества *узлов,* некоторые из которых соединены *звеньями* графа (*каналами*). В работе предлагается общее понятие *логистической сети* как ориентированного графа (кратко орграфа), для моделирования сервиса любого рода, доставляемого по *дугам*, т.е. ориентированным звеньям между узлами. В орграфе каждому звену приписано направление доставки сервиса: от входного узла к выходному.

В логистической сети для каждой дуги указана ее *эффективность*, определяемая как отношение объема чистого полезного сервиса на выходе дуги к объему затраченного сервиса на ее входе (за конкретный временной интервал). В терминах теории графов такая сеть представляет собой дуго-взвешенный орграф, в котором всем дугам $e_k$ приписаны веса $\eta_k$ — действительные числа из промежутка (0;1]. Эти веса моделируют эффективности соответствующих дуг. Логистические сети моделируют ситуации, в которых имеет место передача материальной субстанции (или материальных объектов), энергии, информации или финансов по дугам сети. При доставке сервиса по дуге $e_k$ потери прямо пропорциональны объему сервиса на ее входе с коэффициентом $1-\eta_k$.

Логистическая сеть называется *односторонней*, если между любыми двумя ее узлами $u$ и $v$ сервис доставляется или только из $u$ в $v$, или только из $v$ в $u$, или вообще не доставляется. Сеть называется *двусторонней*, если из того, что сервис доставляется из $u$ в $v$ следует, что он также доставляется из $v$ в $u$, т.е. $u$ и $v$ соединены парой дуг с противоположными направлениями, $uv$ и $vu.$ Сеть, не являющуюся ни односторонней, ни двусторонней, будем называть *смешанной*. У сети любого из этих видов каждую пару противоположных дуг $uv$ и $vu$ будем заменять на одно неориентированное звено с тем же весом, например, так поступаем с парой дуг $ed$ и $de$ в смешанной сети на первой диаграмме рис. 1. Если в двусторонней сети у каждой пары противоположных дуг веса одинаковы, сеть называется *симметричной двусторонней сетью* и преобразуется к неориентированному графу, причем предполагается, что сервис доставляется в обоих направлениях каждого звена с равными эффективностями.

В разделе 3 представлены возможные приложения предлагаемой логистической модели: материальные, энергетические, информационные и финансовые сети. В разделах 4 и 5 решены две задачи (соответственно): I) в данной логистической сети общего вида с выделенной парой конкретных узлов $a$ и $z$ найти цепь из $a$ в $z$, имеющую максимальную эффективность, и II) для данной связной симметричной двусторонней логистической сети найти гарантированный минимальный уровень эффективности сервиса (между любой парой не указанных конкретно узлов).

**2. Понятие эффективности логистической цепи.** *Логистическая цепь* из узла $a$ в узел $z$ — это последовательность дуг, конец каждой из которых, кроме последней, является началом следующей, причем первая дуга выходит из $a$, а у последней дуги цепи только один конец — узел $z$. *Эффективность сервиса на одном звене* логистической сети определяется как отношение объема чистого полезного сервиса на выходе звена к объему затраченного сервиса на его входе:

$$(1) \qquad \eta = \frac{|\text{Service}_{\text{вых.}}|}{|\text{Service}_{\text{вход.}}|}$$

Здесь $\eta$ — безразмерная величина, $0 < \eta \leq 1$. Формула (1) также служит определением общей эффективности сервиса во всей цепи как отношения сервиса на выходе цепи к сервису на ее входе. Нетрудно показать, что общая эффективность сервиса в $q$-звенной цепи вычисляется как произведение эффективностей ее звеньев (см. [6, с. 17]):

$$(2) \qquad \eta_{\text{общ.}} = \eta_1 \eta_2 \cdots \eta_q$$

**3. Основные классы логистических сетей.** Приведем четыре основных класса логистических сетей, второй из которых представлен в [6, с. 16], а третий в [9, с. 4].

3.1. *Материальная логистическая сеть*. Здесь звенья сети представляют транспортировочные магистрали, по которым передается материальная субстанция или материальные объекты, например, зарегистрированный багаж на авиарейсах. Эффективность звена здесь — это отношение количества материальной субстанции (в весовом, объемном или штучном измерениях) на выходе звена к ее количеству на входе. Конкретные примеры: транспортные логистические сети [1], сети централизованного теплоснабжения и сети трубопроводов (жидкости или газа) [6].

Интересный факт отмечен в зарубежных исследованиях [7]: оказывается, в течение 2011 года 26 млн. единиц зарегистрированного багажа во всем мире сбивались с маршрута на международных авиарейсах. Если бы все эти потери случались бы в одном аэропорту размера Международного аэропорта Филадельфии, получилось бы, что в среднем за 2011 год у *каждого* пассажира терялась бы одна единица зарегистрированного багажа!

3.2. *Энергетическая логистическая сеть*. Здесь звенья сети представляют линии передач энергии (обычно линии электропередач). Эффективность звена энергетической цепи определяется как отношение чистой полезной энергии на выходе цепи к затраченной энергии на входе. Важным примером являются электрические энергосети.

3.3. *Информационная логистическая сеть.* Здесь звенья сети представляют каналы связи. Конкретным примером канала связи служит описанный в [9, с. 4] *бинарный симметричный канал*, на вход которого подается бит, причем вероятность того, что бит передается по каналу $e_k$ правильно, имеет определенное значение $p_k$ ($0 < p_k \leq 1$), которое в теории кодирования называется *надежностью* канала. В нашей общей теории логистических сетей мы интерпретируем это значение как эффективность $\eta_k$ ($= p_k$) звена $e_k$. Тогда получается, что $\eta_k$ выражает долю текста, в среднем передаваемую правильно по каналу $e_k$, что согласуется с определением (1).

В случае информационной цепи надо сделать оговорку, что формула (2) верна, если под $|\text{Service}_\text{вых.}|$ понимать объем текста, передаваемого правильно *каждом* звене. Если же контролировать правильность передачи бита, например, в двухзвенной цепи, сравнивая лишь его начальное и самое конечное значения, то две ошибки подряд компенсируют друг друга, и итоговый бит получается таким же, как исходный; в такой постановке если эффективности звеньев равны $\eta_1$ и $\eta_2$, то общая эффективность цепи будет не $\eta_1\eta_2$, а уже $\eta_1\eta_2 + (1-\eta_1)(1-\eta_2)$.

3.4. *Финансовая логистическая сеть.* Здесь звенья сети представляют каналы транзакций. Если банк взимает комиссионный сбор в размере КСБ% от суммы транзакции, то в качестве эффективности такого канала $e_k$ следует взять величину $\eta_k = 1 - \text{КСБ}/100$.

**4. Нахождение максимально эффективной логистической цепи.**

**Задача I.** В данной логистической сети общего вида с выделенной парой узлов $a$ и $z$ найти цепь из $a$ в $z$, имеющую максимальную эффективность.

Сразу же отметим, что цепь из $a$ в $z$ с максимальной суммой весов дуг не обязательно будет иметь их максимальное произведение. Например, из пары двухзвенных цепей из $a$ в $z$ с весами дуг 0,8 и 0,3 у первой цепи, и 0,5 и 0,5 у второй, первая будет иметь бо́льшую сумму весов, а вторая большее произведение. Будем говорить, что первая цепь больше по сложению (т.е. в *аддитивном* смысле), а вторая больше по умножению (т.е. в *мультипликативном* смысле). Однако можно утверждать (см. утверждение 1 ниже), что цепь из $a$ в $z$ с минимальной суммой весов Хартли-Шеннона уже будет цепью с максимальным произведением исходных весов дуг.

Для решения задачи I мы вместо $\eta_k$ вводим новые веса дуг, которые будем называть *весами Хартли-Шеннона,* следующим образом: $\iota_k = |\log_b(\eta_k)| = -\log_b(\eta_k)$, где $b$ — параметр ($b > 1$). В нашем исследовании важную роль играет изоморфизм между аддитивной полугруппой действительных чисел $y \in [0; +\infty)$ с нейтральным элементом 0 и мультипликативной полугруппой действительных чисел $x \in (0;1]$ с нейтральным элементом 1: если $y_1 = |\log_b(x_1)|$ и $y_2 = |\log_b(x_2)|$, то

(3) $\qquad y_1 + y_2 = |\log_b(x_1 x_2)| \quad$ или $\quad \iota_1 + \iota_2 = |\log_b(\eta_1 \eta_2)|$.

Таким образом, произведению весов $\eta_k$ дуг соответствует сумма их весов Хартли-Шеннона $\iota_k$. В то время как $\eta_k$ выражает эффективность дуги, $\iota_k$ выражает то, что мы предлагаем называть ее *убыточностью*. Значит, *умножению эффективностей дуг соответствует сложение их убыточностей,* и общая убыточность логистической цепи равна сумме убыточностей ее звеньев. Наиболее эффективная цепь наименее убыточна.

**Утверждение 1.** Из логистических цепей из узла $a$ в узел $z$ произведение весов (дуг) $\eta_k$ максимально у той, у которой минимальна сумма весов Хартли-Шеннона $\iota_k$.

В силу (3), утверждение 1 следует из того, что при $b > 1$ функция $y = |\log_b(x)|$ монотонно убывает на промежутке $(0;1]$. Хотя в нашей теории непринципиально, по какому основанию $b > 1$ брать логарифмы, в случае информационной сети естественнее положить $b = 2$, т.е. $\iota_k = |\log_2(\eta_k)| = -\log_2(\eta_k) = \log_2(1/\eta_k)$. Поскольку $\eta_k$ выражает долю текста, в среднем передаваемую правильно по каналу $e_k$, получается, что в среднем передается правильно один бит из $1/\eta_k$ битов. Минимальное количество битов информации, достаточное для выявления этого правильного бита в точности равно весу Хартли-Шеннона $\iota_k$. Последнее утверждение есть известная формула Хартли [8], обобщенная Шенноном [10].

Мы предлагаем следующий алгоритм решения задачи I. Он верен в силу утверждения 1.

**Алгоритм 1.** *Вход:* Логистическая сеть общего вида с выделенной парой узлов $a$ и $z$. *Подготовительная часть:* Заменить веса дуг на соответствующие веса Хартли-Шеннона. *Основная часть:* Найти цепь из *a* в *z* с минимальной суммой весов Хартли-Шеннона. *Выход:* Найденная цепь максимально эффективна.

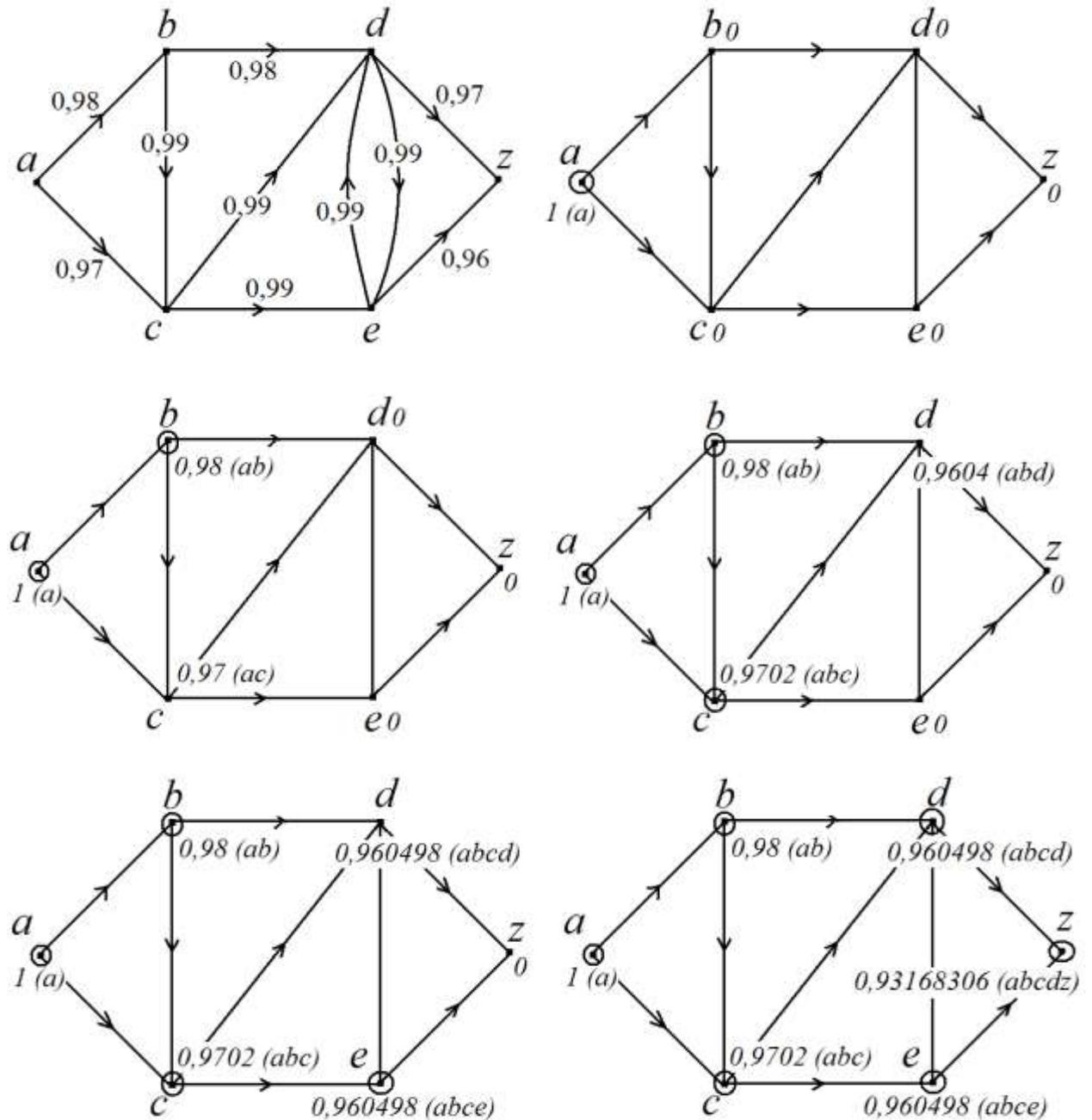

Рис. 1. Решение задачи I при помощи алгоритма 2.

Для выполнения основной части можно использовать известный алгоритм Дейкстры [3, с. 13], [5, с. 189]. Изоморфизм (3) вкупе с утверждением 1 и со свойством монотонности функции $y = |\log_b(x)|$ на (0;1] позволяет приспособить алгоритм 1 в терминах исходных весов без перехода к весам Хартли-Шеннона. Для этого надо заменить "0" на "1",

"∞" на "0", "min" на "max", а также изменить все знаки неравенств на противоположные. В результате получается мультипликативный аналог алгоритма Дейкстры.

Описанию алгоритма предпошлем несколько пояснений. Хотя в данной логистической сети весá имеются только у дуг, в процессе работы алгоритма узлам тоже будут назначаться весá (по определенным правилам). Кроме этого, напротив каждого узла $u$ в процессе работы алгоритма будет записываться определенная последовательность узлов (в скобках после указания веса), которую мы будем называть "историей" узла $u$. Когда вес узла пересчитан окончательно, узел обводится в кружок. Введем еще два основных шага, которые в процессе работы алгоритма будут чередоваться.

*1-й основной шаг:* Из узлов, еще не обведенных в кружок, выбрать тот, у которого наибольший текущий вес и обвести его в кружок. Если таких узлов более одного, обвести в кружок любой один из них.

*2-й основной шаг:* Для узла $v$, только что обведенного в кружок, попытаться обновить вес каждого еще не обведенного в кружок узла $u$, в который есть дуга $vu$ из $v$, по следующему правилу: если произведение веса узла $v$ на вес дуги $vu$ больше текущего веса узла $u$, то заменяем вес узла $u$ на это произведение, иначе вес узла $u$ не обновляется. В случае фактического обновления веса узла $u$, текущая история узла $u$ заменяется на текущую историю узла $v$, к которой справа приписывается символ $u$.

**Алгоритм 2 (мультипликативный аналог алгоритма Дейкстры).** *Вход:* Логистическая сеть общего вида с выделенной парой узлов $a$ и $z$. *Подготовительная часть:* Узлу $a$ присваивается вес 1, а всем остальным узлам вес 0. Также узлу $a$ присваивается история, состоящая из одного символа $a$. *Основная часть:* К получающейся сети применяются последовательно в чередующемся порядке 1-й и 2-й основные шаги, начиная с 1-го. Процесс завершается, как только узел $z$ оказывается обведенным в кружок. Процесс также завершается, когда после очередного выполнения 2-го основного шага узел $z$ остается необведенным, но все необведенные узлы имеют текущий нулевой вес (такая ситуация возникает, когда вообще не существует логистической цепи из $a$ в $z$). *Выход:* История узла $z$ показывает максимально эффективную цепь из $a$ в $z$, а вес узла $z$ показывает эффективность этой цепи.

В силу (3) и монотонности log очередность обведения узлов в кружок в алгоритме 2 такая же, как в алгоритме 1. Значит, алгоритм 2 верен, потому что алгоритм 1 верен.

**Пример 1.** От подстанции *a* энергия доставляется в пункты *b*, *c*, *d*, *e* и *z* по смешанной энергетической сети, показанной на 1-й диаграмме рис. 1. Для каждой дуги указана ее эффективность. Для подачи энергии от *a* в *z* найти цепь максимальной энергоэффективности.

Для решения примера 1 используем алгоритм 2. Работа этого алгоритма иллюстрируется диаграммами на рис. 1. На последней диаграмме порядок обведения узлов в кружок следующий: сначала *d* (после пересчета весов узлов *d* и *z* из узла *e*), затем *z* (после пересчета веса узла *z* из узла *d*). Ответ к задаче находится на последней диаграмме напротив узла *z*, а именно: *abcdz* — самая эффективная цепь, ее энергоэффективность равна 0,93168306.

Заметим, что в примере 1 в процессе работы алгоритма возникает паритет весов у узлов *d* и *e* (см. предпоследнюю диаграмму на рис. 1). Поэтому альтернативный вариант работы алгоритма состоит в том, что на предпоследней диаграмме в кружок обводится сначала узел *d*, затем *e* (потому что у *e* больший вес, чем текущий вес узла *z*) и, наконец, *z* (после пересчета веса узла *z* из *e* без его обновления). Ответ получается такой же.

**5. Нахождение гарантированного минимума эффективности сервиса в сети.**

**Задача II.** В данной связной симметричной двусторонней логистической сети найти гарантированный минимальный уровень эффективности сервиса между любой парой не указанных конкретно узлов, т.е. такой уровень $\eta_{\text{гар.min}}$, что из любого узла в любой другой найдется цепь с эффективностью, не меньшей $\eta_{\text{гар.min}}$. В качестве тривиального решения можно взять нулевой уровень, но мы заинтересованы в нахождении максимально возможного уровня для $\eta_{\text{гар.min}}$.

*Остовным деревом* в данном неориентированном графе называется связный подграф, не содержащий циклов и покрывающий все узлы данного графа. Как обосновано в [3, с. 88], если веса всех звеньев графа положительны, остовное дерево с минимальной суммой весов имеет и минимальное их произведение. Доказательство [3, с. 88] можно слегка модифицировать и доказать следующее утверждение.

**Утверждение 2.** Если веса всех звеньев графа положительны, остовное дерево с максимальной суммой весов звеньев имеет и максимальное их произведение.

В силу утверждения 2 задача II решается адаптацией алгоритма Краскала (изложенного, например, в [2, с. 309], [3, с. 24] и [5]), который отыскивает остовное дерево с максимальной суммой весов звеньев.

**Алгоритм 3.** *Вход:* Связная симметричная двусторонняя логистическая сеть. *Подготовительная часть:* Текущее множество звеньев устанавливается пустым. *Основная часть:* Последовательно делается следующий основной шаг: из звеньев, которые не принадлежат текущему множеству, и добавление которых к текущему множеству не создает в нем цикла, выбирается звено максимально возможного веса и добавляется к текущему множеству. Когда число звеньев в текущем множестве становится на единицу меньше, чем общее число узлов в сети, процесс завершается. *Выход:* Текущее множество звеньев определяет остовное дерево с наибольшим произведением весов звеньев $\Pi_{max}$.

Поскольку эффективность каждого звена логистической сети есть величина между нулем и единицей, из формулы (2) следует, что любая цепь в найденном остовном дереве имеет эффективность, не меньшую $\eta_{\text{гар.min}} = \Pi_{max}$. В качестве цепи соединяющей конкретную пару узлов сети с эффективностью, не меньшей $\eta_{\text{гар.min}}$, достаточно взять цепь в найденном остовном дереве, соединяющую эти узлы.

**Алгоритм 4.** *Вход:* Такой же, как для алгоритма 3. Перебираются все пары узлов сети, и для каждой пары при помощи алгоритма 2 ищется максимальная эффективность сервиса между ними. *Выход:* Наименьшая из найденных эффективностей и будет максимальным уровнем $\eta_{\text{гар.min}}$.

Отметим, что алгоритм 4 также применим к произвольной логистической сети общего вида, причем всегда дает на выходе максимально возможное значение для $\eta_{\text{гар.min}}$, в то время как алгоритм 3 применим только для связных симметричных односторонних сетей и дает значение, близкое к максимальному лишь для достаточно разреженных сетей. Однако алгоритм 4 имеет бо́льшую вычислительную сложность — время его работы по порядку не менее чем в $n^2/\log n$ раз больше времени работы алгоритма 3 даже для густых сетей, где $n$ — число узлов в сети.

**Заключение.** Хотя алгоритм 2 представляется полезным для определения наиболее эффективной цепи доставки сервиса (электроэнергии в примере 1), в расчетах логистических сетей легче оперировать с убыточностями сервиса, как в алгоритме 1, вместо его эф-фективностей, как в алгоритме 2. Причина в том, что произведению эффективностей соответствует сумма убыточностей.

В заключение подчеркнем, что предложенная концепция логистической сети универсальна. В работе показано, как она может служить для моделирования четырех классов сетей: материальных, энергетических, информационных и финансовых. Однако области применимости концепции шире этих четырех классов. Алгоритмы 1 и 2 также могут найти применение в анализе социологических сетей, например, когда пользователь $a$ ищет цепь поручителей, чтобы с максимальной вероятностью добиться принятия "узлом" $z$ нужного для $a$ решения. «Так порука складывалась в цепь поручителей, охватывавшую весь служилый уезд.» — писал еще В.О. Ключевский [4, с. 257].

**Литература**